\input{aipcheck}

\documentclass[
    ,final            
  ]
  {aipproc}

\layoutstyle{6x9}

\begin{document}

\title{Ultraluminous X-ray Sources forming in low metallicity natal environments}


\author{L. Zampieri}{
  address={INAF-Astronomical Observatory of Padova, Vicolo dell'Osservatorio 5, I-35122 Padova, Italy}
}

\author{M. Colpi}{
  address={Università Milano Bicocca, Dipartimento di Fisica G.Occhialini, Piazza delle Scienze 3, I-20126, Milano, Italy}
}

\author{M. Mapelli}{
  address={(Institute for Theoretical Physics, University of Zürich, Winterthurerstrasse 190, CH-8057 Zürich, Switzerland}
}

\author{A. Patruno}{
  address={Astronomical Institute `A. Pannekoek', Univeristy of Amsterdam, Science Park 904, Amsterdam, The Netherlands}
}

\author{T. P. Roberts}{
  address={Department of Physics, Durham University, South Road, Durham DH1 3LE, UK}
}

\begin{abstract}
In the last few years multiwavelength observations have boosted 
our understanding of Ultraluminous X-ray Sources (ULXs). Yet, the most fundamental questions on ULXs 
still remain to be definitively answered: do they contain stellar or intermediate mass black holes? 
How do they form? We investigate the possibility that the black holes hosted in ULXs 
originate from massive (40-120 $M_\odot$) stars in low metallicity natal environments. Such black holes
have a typical mass in the range $\sim 30-90 M_\odot$ and may account for the properties
of bright (above $\sim 10^{40}$ erg s$^{-1}$) ULXs. More than $\sim 10^5$ massive black holes might have 
been generated in this way 
in the metal poor Cartwheel galaxy during the last $10^7$ years and might power most of the ULXs 
observed in it. Support to our interpretation comes from NGC 1313 X-2, the first ULX with a tentative 
identification of the orbital period in the optical band, for which binary evolution calculations show 
that the system is most likely made by a massive donor dumping matter on a $50-100 M_\odot$ black hole. 
\end{abstract}

\maketitle


\section{Introduction}

When, at the beginning of the 80s, point-like, off-nuclear Xray
sources were first detected in the field of nearby galaxies
(see, e.g., \cite{b19}), it was immediately recognised
that the luminosity of a subset of these objects was unusually
large. If physically associated with their host galaxies,
these UltraLuminous X-ray sources (ULXs) had an isotropic luminosity 
well in excess of the Eddington limit for spherical accretion onto a
$10 M_\odot$ compact object. 
Thanks to the unprecedented capabilities offered by some of the major 
X-ray satellites (XMM, Chandra) and optical facilities
(VLT, HST), nowadays more than 150 candidate ULXs have
been detected and many of them have been studied in detail. 
Several pieces of observational evidence
strongly suggest that a large fraction of these sources are accreting 
black hole X-ray binaries with massive donors (see, e.g., \cite{zr09}).

The critical issue is then understanding what is responsible
for the exceptionally high (isotropic) luminosity
of these sources. Two main scenarios have been proposed.
Firstly ULXs could be relatively normal stellar-mass ($\le 20 M_\odot$)
black holes (BHs) that are either anisotropically emitting X-ray binaries
in a peculiar evolutionary stage \citep{b32}, or are truly emitting 
above the Eddington limit via a massive, modified accretion disc structure
(e.g. photon bubble dominated discs \citep{Begel06}; two-phase
super-Eddington, radiatively efficient discs \citep{sd06}; slim discs \citep{b15a}), 
or perhaps via some combination of the two \citep{pout07,King08}.
Secondly, the compact object could simply be bigger,
and the accretion would be in the usual sub-Eddington
regime. In this case the compact object would be an intermediate
mass black hole (IMBH) with a mass in excess of $100 M_\odot$ (e.g. \citep{b11}).


Recently, \cite{zr09} presented a critical revaluation of the available 
observational evidence concerning the BH masses in ULXs, suggesting that
BHs of several hundreds to thousands $M_\odot$ are not required for 
the majority of ULXs. At the same time, models with stellar mass BHs may work for a large fraction
of the ULX population, if the accretion flow has some degreee of beaming and is 
super-Eddington (e.g. \cite{King09}), but rather extreme conditions are needed to account 
for ULXs above a luminosity $\sim 10^{40}$ erg s$^{-1}$.
Here we highlight an alternative scenario
in which a proportion of ULXs contain $\sim 30- 90 M_\odot$ BHs formed in a low 
metallicity environment and accreting in a slightly critical regime.

\section{A different interpretation}

In our scenario bright ULXs may contain BHs with masses above 30-40 $M_\odot$ and 
up to $\sim 80-90 M_\odot$, formed from ordinary stellar evolution of massive 
($40-120 M_\odot$) stars in a low metallicity natal environment. While this idea has
already been suggested before (e.g. \cite{b55,b10a,b73}), it has
not yet been explored quantitatively in detail.
In stars with main sequence masses above $\sim 25-30 M_\odot$,
at the time of iron core collapse, the early accretion of the inner mantle 
before shock passage and the fallback of material afterwards
cause the newly formed proto-neutron star to collapse to a BH after
the supernova explosion.
At solar metallicity, these fallback BHs reach at most $\sim 10 M_\odot$
as the stellar envelope is effectively removed through line-driven
winds. For sub-solar metallicities, however, this mechanism becomes
progressively less efficient and stars with masses above $\sim 30-40
M_\odot$ may retain rather massive envelopes at the time of
explosion. The supernova shock wave then loses more and more energy in
trying to unbind the envelope until it stalls and most of the star
collapses to form a BH (direct BH formation) with a mass comparable to that of the
pre-supernova star \citep{fr99}. These may be the BHs hosted in some
ULXs. Their mass would not be significantly larger than $\sim 80-90
M_\odot$ as above $\sim 100-120 M_\odot$ a star undergoes pulsational
pair-instability in its core and most of the envelope mass is expelled.

According to the adopted mass loss history, the final mass of a massive star 
may differ up to a  factor of $\sim 2$, or even more for clumpy winds.
Additional uncertainty is caused by the dependence of mass loss
on metallicity. A scaling law $\propto Z^{0.5}$ is often adopted for
hot stars (see, e.g., \cite{nl00}). Considering a star with an initial
mass of $100 M_\odot$, its final mass may be in the interval $\sim 3-6 
M_\odot$ for $Z\approx Z_\odot$ and $\sim 30-70 M_\odot$ for 
$Z\approx 0.1 Z_\odot$ \cite{zr09}.

Owing to their larger final masses, the fate
of stars with sub-solar metallicity is likely to be be quite different
from that of higher metallicity stars. Although different authors
obtain different results for the mass of the compact remnant, it is
not unreasonable to think that, if an envelope more massive than $\sim
30-40 M_\odot$ is retained at the time of explosion, a low metallicity
($Z\approx 0.1 Z_\odot$) star may collapse directly to form a BH of
comparable mass. 
Significant stellar rotation (hundreds of km s$^{-1}$) may change this picture
somewhat. However, if the core is not rapidly rotating, there is no good 
reason why most of the star should not collapse into a BH.
At variance with IMBHs, the formation of these very
massive stellar remnant BHs does not require an exotic, new mechanism
but is referable to ordinary stellar evolution. At the same time, only
modest beaming ($\sim 0.5$) or slight violations of the Eddington
limit (a factor of a few) would be needed to account for the
luminosity of bright (above $\sim 10^{40}$ erg s$^{-1}$) ULXs.

\section{Testing the massive BH interpretation}

A crucial aspect of the interpretation of ULXs in terms of BHs from
the direct collapse of low-$Z$, massive stars is the metallicity of
the environment in which ULX binaries form. The scarce measurements available
and the discrepancy between optical and X-ray data prevent at this stage
to reach a definitive conclusion. In our proposed scenario ULXs
should show some evidence of correlation (in terms of position and
average luminosity) with low metallicity environments. So, one of the
definitive tests of our proposal would be to survey ULX locations, and
determine whether a relationship between ULX luminosity and local
metallicity was evident in a large enough sample to provide
statistically meaningful results. It is worth noting that, recently, \cite{prest07} and \cite{sf08}
succeeded in performing dynamical mass measurements using Gemini and
Keck spectra of the Wolf-Rayet optical counterpart of IC 10 X-1, a
variable X-ray source in the the Local Group metal poor starbust
galaxy IC 10. They find a BH mass in the range $23-33 M_\odot$, which
represents the most massive BH known to exist in a binary system and
definitely corroborates our interpretation.


A crucial benchmark to test our interpretation is provided by the
Cartwheel galaxy. This has a rather low metallicity ($Z \sim 0.05 Z_\odot$, 
measured in the nebulae of the outer ring which are forming stars right now
\cite{fw77}) and hosts a large number of ULXs ($\sim 17$, \cite{wt04}).
We estimated the number of massive ($\ge 40 M_\odot$) BHs $N_{BH}$
produced during a burst of star formation, assuming that they are distributed
according to the stellar IMF \cite{map09}. For a star formation rate of
$\sim 20 M_\odot$ yr$^{-1}$ \cite{may05} and a duration of the star burst 
of $\sim 10^7$ yr, we find $N_{BH}\sim 10^5$ (with a slight dependence on the
adopted IMF). The total mass ended up in massive BHs turn out to 
be $M_{BH} \sim 10^7 M_\odot$, corresponding to $\sim$ 5\% of the total 
stellar mass in the ring produced during the burst.
Also, the production efficiency of ULXs ($N_{ULXs}/N_{BH}$)
is estimated to be $\sim 10^{-4}$, which appears reasonable if compared to 
that obtained from independent estimates (from dynamical and/or binary evolution 
models \cite{map09}).


Finally, independent evidence in support of our interpretation may come also
from NGC 1313 X-2, one of the most studied ULXs to date, located in a low
metallicity environment ($Z\simeq 0.004-0.008$; e.g. \cite{pz09} and references
therein). Recently, \cite{liu09} tentatively identified a modulation 
of $6.12\pm 0.16$ d in the $B$ band HST lightcurve of this ULX. They interpreted
the modulation as the orbital period of the system.
Assuming that this identification is correct, we used all the optical
data available for NGC 1313 X-2 and compared them with the evolution of an ensemble of 
irradiated X-ray binary models in order to constrain the nature of its compact 
accretor \cite{pz09}. We restricted the
candidate binary system to be either a $\sim 50-100
M_\odot$ BH accreting from a $15 M_\odot$ main sequence star or a
$20 M_\odot$ BH with a $12-15 M_\odot$ giant donor. If the
modulation of $\sim 6$ days is confirmed, a stellar-mass BH model
becomes unlikely and we are left with the only possibility that the
compact accretor in NGC 1313 X-2 is a massive BH of $\sim
50-100 M_\odot$, in agreement with the interpretation that it may contain a massive BH.

\section{Conclusions}


We investigated in detail an alternative scenario in which bright ULXs
contain BHs with masses above $\sim 30-40 M_\odot$ and up to $\sim
80-90 M_\odot$, produced by stars with initial, main sequence mass
above $\sim 40 M_\odot$. The formation of these very massive stellar remnant BHs does not
require an exotic, new mechanism but is referable to ordinary stellar
evolution. For luminosities above $\sim 10^{40}$ erg s$^{-1}$, this would
imply only modest violations of the Eddington limit, attainable
through very modest beaming and/or slightly super-critical accretion.
Measurements of the metallicity of the ULX environment and surveys 
of ULX locations looking for a statistically meaningful relationship 
between position, average luminosity and local metallicity will 
provide a definitive test of our proposal.

\begin{theacknowledgments}
LZ and MC acknowledge financial support from INAF through grant PRIN-2007-26.
\end{theacknowledgments}


\end{document}